\documentclass[
aps,
prb,
showpacs,
floatfix,
reprint,
twoside,
]
{revtex4-1}
\usepackage{amssymb}

\usepackage{amsmath}
\usepackage[T1]{fontenc}
\usepackage{fourier}
\usepackage{graphicx}
\usepackage[colorlinks=true,
allcolors=blue,
]{hyperref}

\begin{document}

\title{Time-dependent equation for the magnetic order parameter\texorpdfstring{\\}{}
near the quantum critical point in multiband superconductors with a spin density wave}
\author{Andreas Moor$^1$, Anatoly F.~Volkov$^{1,2}$ and Konstantin B.~Efetov$^1$}
\affiliation{$^1$Theoretische Physik III, Ruhr-Universit\"{a}t Bochum, D-44780 Bochum, Germany\\
$^2$Donostia International Physics Center (DIPC), Manuel de Lardizabal 5, E-20018 San Sebastian, Spain}

\begin{abstract}
Using a simple two-band model for Fe\nobreakdash-based pnictides and the generalized Eilenberger equation, we present a microscopic derivation of a time-dependent equation for the amplitude of the spin density wave near the quantum critical point where it turns to zero. This equation describes the dynamics of the magnetic\nobreakdash---$m$, as well as the superconducting order parameter\nobreakdash---$\Delta$. It is valid at low temperatures~$T$ and small~$m$ (${T, m \ll \Delta}$) in a region of coexistence of both order parameters,~$m$ and~$\Delta$. The boundary of this region is found in the space of the nesting parameter~$\{\mu_{0},\mu_{\phi}\}$ where~$\mu_{0}$ describes the relative position of the electron and the hole pockets on the energy scale, and~$\mu_{\phi}$ accounts for the ellipticity of the electron pocket. At low~$T$ the number of quasiparticles is small due to the presence of the energy gap~$\Delta$, and therefore the quasiparticles do not play a role in the relaxation of~$m$. This circumstance allows one to derive the time-dependent equation for~$m$ in contrast to the case of conventional superconductors for which the time-dependent Ginzburg--Landau equation can be derived near~$T_{\text{c}}$ only in some special cases (high concentration of paramagnetic impurities\cite{G-Eliash}). In the stationary case the derived equation is valid at arbitrary temperatures. We find a solution of the stationary equation which describes a domain wall in the magnetic structure. In the center of the domain wall the superconducting order parameter has a maximum, which means a local enhancement of superconductivity. Using the derived time-dependent equation for~$m$, we investgate also the stability of a uniform commensurate SDW and obtain the values of~$\{\mu_{0}, \mu_{\phi}\}$ at which the first order transition into the state with~${m = 0}$ takes place or the transition to the state with an inhomogeneous SDW occurs.
\end{abstract}

\date{\today}
\maketitle

\section{Introduction}

Over the last decade, the interest in the quantum phase transitions~(QPT) increased, i.e., in transitions of a system from one state to another, that may occur if a parameter of the system is varied by, e.g., doping at zero temperature (for a review see Refs.~\onlinecite{Sachdev,Balatsky10} and references therein). These transitions can take place in systems with two competing order parameters~(OP). An example are the so-called Fe\nobreakdash-based pnictides, where, as it was established theoretically and experimentally,\cite{Norman,Mazin,Wilson,Stewart,Chubukov,Hirschfeld_Korshunov_Mazin_11} superconductivity~(SC) coexists with the spin density wave~(SDW), see also Ref.~\onlinecite{Shibauchi_et_al} and references within. As is generally known, superconductivity is characterized by the superconducting OP~$\Delta$, whereas the spin density wave---by the antiferromagnetic ordering OP~$\mathbf{m}$, i.e., by the magnetic moment in one of the magnetic sublattices.

Varying the doping level~$x$ in these superconductors, the amplitude~$m$ of the SDW OP can become zero at a finite value of~$\Delta$ and a certain point~$x_\text{m}$ as well as the SC OP~$\Delta$ can vanish at a finite value of~$m$ and a certain~$x_\text{s}$. At low temperatures, these points can be called quantum critical points~(QCP).

A sharp peak in the doping dependence of the London penetration length~$\lambda_{\text{L}}$ has been observed in a recent work.\cite{Hashimoto_Science} This fact was conjectured as a sign of the strong quantum fluctuations near the QCP. Based on the diagram technique for the Green's functions the contribution of the fluctuations of~$m$ to the response of the superconductor to the magnetic field was calculated and it was shown that, indeed, $\lambda_{\text{L}}$~has a peak near~$x_\text{m}$.\cite{Levchenko_Vavilov_Khodas_Chubukov_13,Nomoto_Ikeda_arXiv_13,Chowdhury_et_al_arXiv_13}

The properties of superconductors can be completely described in terms of the Green's functions~$G(\mathbf{r},t; \mathbf{r}^{\prime},t^{\prime})$. The equations for these functions, which in case of superconductors consist of normal and anomalous Green's functions, have been derived by Gor'kov.\cite{Gor'kov} In most cases, the information contained in~$G(\mathbf{r},t; \mathbf{r}^{\prime },t^{\prime })$ is not needed, and it is sufficient to know the Green's functions at coinciding points in space ${\mathbf{r}=\mathbf{r}^{\prime}}$. In this approximation, the quasiclassical Green's functions~$g(\mathbf{r},t; t^{\prime})$ introduced by Eilenberger obey the Eilenberger equations that have been derived on the basis of the Gor'kov's equations.\cite{Eilenberger,*Larkin_Ovchinnikov_69,LO,Serene_Rainer_83} They are simpler than the Gor'kov's equations because the functions~$g(\mathbf{r},t; t^{\prime})$ depend only on one coordinate~$\mathbf{r}$ and on the orientation of the momentum~$\mathbf{p}$. The method of quasiclassical Green's functions has been well developed and used successfully to describe many properties of superconductors (for a review, see Refs.~\onlinecite{RammerSmith,BelzigRev,Kopnin}).

However, the simplest tool for studying superconductivity are the Ginzburg--Landau (G--L) equations which have been derived\cite{GL} on a phenomenological basis even before the microscopic theory of superconductivity was developed by Bardeen, Cooper and Schrieffer in~1957.\cite{BCS} The derivation of the G--L equations is based on the Landau's approach to the description of the phase transition of the second type in the vicinity of the critical point~$T_{\text{c}}$.\cite{L-L5} As is well known, in this method the thermodynamic potential~$\Omega$ is expanded in powers of the amplitude~$\eta$ of the~OP which is small near~$T_{\text{c}}$. The dynamics of~$\eta$ is described by a simple equation
\begin{equation}
\partial_t \eta = -\gamma_{\text{r}} \partial_\eta \Omega
\label{eqn:Eta}
\end{equation}
where~$\gamma_{\text{r}}$ is a phenomenological relaxation rate.\cite{L-L10} Generally speaking, such a simple equation is not valid in the case of superconductors. As was shown by Gor'kov and Eliashberg, a simple generalization of the G--L equations for a time-dependent~$\Delta$ is possible only in a rather exotic case of superconductors with a high concentration of paramagnetic impurities when the gap in the excitation spectrum disappears.\cite{G-Eliash} In ordinary superconductors, the dynamics of~$\Delta$ is determined by the time evolution of the quasiparticle distribution function~$f_{\text{qp}}$. This means that the kinetic equation for~$f_{\text{qp}}$ should be solved with account for inelastic relaxation processes.\cite{G-Eliash,Kopnin}

In this paper, we derive an equation for~$m$ similar to~Eq.~(\ref{eqn:Eta}) in the vicinity of the QCP~$x_{\text{m}}$, when the inequality ${m \ll \Delta}$ holds. This equation describes both spatial and temporal behavior of the amplitude of the SDW~$m$. We consider the case of low temperatures, ${T \ll \Delta}$, when the number of quasiparticles is small (at small~$m$ the gap in the excitation spectrum is~$\Delta$) and, therefore, they do not affect essentially the OPs~$\Delta$ and~$m$. This is the essential difference from the case of ordinary superconductors in which the derivation of the Ginzburg--Landau equation is justified, strictly speaking, only near the superconducting critical temperature~$T_{\text{c}}$ where the number of quasiparticles is large and they determine the dynamics of the superconducting order parameter~$\Delta$. In particular, we find the region in the space of nesting parameters where the derived equation is valid and write down a solution for~$m(x)$ that describes a domain wall~(DW).

We use the mean-field approximation and do not take into account fluctuations of the order parameters. This means that our considerations are valid not too close to the QCP (see Refs.~\onlinecite{Nomoto_Ikeda_arXiv_13,Chowdhury_et_al_arXiv_13}). In addition, we restrict ourselves to the case of a fixed magnetization direction, i.e., the derived equations do not take into account a possible rotation of the magnetization~$\mathbf{m}$. The plan of the paper is as follows.

In Sec.~\ref{sec:EilenbergerEquation} we write the generalized Eilenberger equations for a rather general case of a non-ideal nesting when the superconducting and magnetic OPs coexist.\cite{Moor11,*Moor12,*Moor13} The expressions for the quasiclassical Green's functions for this case are presented in Appendix~\ref{sec:Coefficients}. Using the generalized Eilenberger equation,  we derive in Sec.~\ref{sec:GL-like_Eq} the stationary, Eq.~(\ref{eqn:TimeIndependentEquationForM}), and in Sec.~\ref{sec:GL-like_Eq_timedependent}---the time-dependent equation, Eq.~(\ref{eqn:TimeDependentEquationForM}), for the amplitude of the SDW. This equation is similar to Eq.~(\ref{eqn:Eta}) and to the time-dependent Ginzburg--Landau equation. We find the region in the space of the nesting parameter (represented by~$\mu_{0}$ and~$\mu_{\phi}$), where the state with a uniform SDW (${m \neq 0}$) is stable, and determine the values~$\mu_{0}$ and~$\mu_{\phi}$ at which the first order transition into the state with ${m = 0}$ or a transition into a state with non-uniform SDW occurs. The parameter~$\mu_{0}$ describes the relative position of the electron and the hole bands on the energy scale, and~$\mu_{\phi }$ accounts for the ellipticity of the electron pocket. In Sec.~\ref{sec:DomainWall} we present solutions for~$m(x)$ and~$\Delta(x)$ that describe a domain wall and, respectively, the enhancement of superconductivity approaching the center of the domain wall. The obtained results are discussed in the concluding Section~\ref{sec:Discussion}.

\section{Eilenberger Equation}
\label{sec:EilenbergerEquation}

As in our previous works,\cite{Moor11,*Moor12,*Moor13} we consider a simple two-band model of a superconductor with an SDW developed in Refs.~\onlinecite{Chubukov09,Schmalian10}. This model of a multiband (hole and electron bands) superconductor allows for the coexistence of the superconducting and the magnetic, i.e., SDW order parameters. It has been shown that the most energetically favorable state in this model is the so-called s$_{+-}$\nobreakdash-state, in which the superconducting order parameters in the electron and hole bands have opposite signs.\cite{Mazin_Singh_Johannes_Du_08,Chubukov09,Schmalian10,Kuroki_et_al_08} This widely accepted model provides a good description of the coexistence realm and the qualitative behavior of the SDW, but fails in describing the correct doping dependence of the superconducting order parameter outside the coexistence region yielding a doping-independent~$\Delta$ as opposed to experimentally measured superconducting domes in real materials like Fe\nobreakdash-based pnictides (see, e.g., Refs.~\onlinecite{Nandi_et_al_10,Mazin2010}).

It is well known that the method of quasiclassical Green's functions is a powerful and effective method for the theoretical study of superconductors.\cite{Eilenberger,LO,RammerSmith,BelzigRev,Kopnin} This method has been applied also to systems with different kind of order parameters. For example, the systems with the charge (spin) density waves may be described by the quasiclassical Green's functions.\cite{ArtVolkovCDW,Moor11} Recently, this method was applied to describe topological superconductors.\cite{Nagai_Nakamura_Machida_arXiv_13} In several papers the method of quasiclassical Green's functions has been applied for the description of two-band superconductors.\cite{GurevichVinokur03,*Gurevich03,KoshelevGolubov03,Anish07,SilaevBabaev12} In the considered case of multiband superconductors with the SDW these functions are $8\times8$~matrices $\check{g}$ in the band, Gor'kov--Nambu and spin spaces. These matrix functions obey a generalized Eilenberger equation\cite{Moor11} that contains the superconducting and antiferromagnetic order parameters
\begin{equation}
\big(\check{\mathrm{X}}_{030} \partial_t \check{g} + \partial_{t^{\prime}} \check{g} \check{\mathrm{X}}_{030}\big) + \mathbf{v} \mathbf{\nabla} \check{g} + \mathrm{i} \big(\check{\Lambda} \check{g} - \check{g} \check{\Lambda}\big) = 0 \,,
\label{eqn:Eilenberger_general}
\end{equation}
where~${\mathbf{v} = \partial_{\mathbf{p}} \xi(\mathbf{p}_{\text{F}})}$ is the vector of Fermi velocity and the matrix ${\check{\Lambda} = \mu \check{\mathrm{X}}_{300} - \mathrm{i} \big( \Delta^{\prime} \check{\mathrm{X}}_{323} + \Delta^{\prime \prime} \check{\mathrm{X}}_{013} \big) + \mathrm{i} m \check{\mathrm{X}}_{123} }$ consists of the parts describing the deviation from the ideal nesting---${\mu \equiv \mu_{\mathbf{p}} = \big(\xi_1(\mathbf{p}) + \xi_2(\mathbf{p})\big) / 2 = \mu_{0} + \mu_{\phi} \cos(2\phi)}$, the superconductivity---${\Delta = \Delta^{\prime} + \mathrm{i} \Delta^{\prime \prime}}$, and the spin density wave---$m$, the magnetization vector being oriented along the $z$\nobreakdash-direction. The SC order parameter is related to the SC order parameters in the first and the second band as ${\Delta_1 = \Delta^* = -\Delta_2}$ in the case of the~s$_{+-}$\nobreakdash-pairing (we assume, for simplicity, ${|\Delta_1| = |\Delta_2| = |\Delta|}$). In the expression for~$\mu$ the term~$\mu_0$ takes into account the difference in the electron and hole masses, whereas~$\mu_{\phi}$ describes the ellipticity of the electron pocket~(${m_{2,x} \neq m_{2,y}}$), see Appendix~\ref{app:QCGF}. The matrices ${\check{\mathrm{X}}_{ijk} = \hat{\rho}_{i} \cdot \hat{\tau}_{j} \cdot \hat{\sigma}_{k}}$ are composed of Pauli matrices in the band~($\hat{\rho}$), Gor'kov--Nambu~($\hat{\tau}$) and spin space~($\hat{\sigma}$), respectively, with additional $2\times2$~unity matrices~($\hat{\rho}_0$, $\hat{\tau}_0$ and $\hat{\sigma}_0$), and the dot denotes the tensor product. Note that the coexistence of the superconducting and the magnetic order parameters is only possible\cite{Chubukov09,Schmalian10} if ${\mu \neq 0}$.

The Eilenberger equation~(\ref{eqn:Eilenberger_general}) for the quasiclassical Green's function matrix~$\check{g}$ is supplemented with the normalization condition
\begin{equation}
\check{g} \cdot \check{g} = \check{1} \,. \label{eqn:NormCondition}
\end{equation}

The superconducting and magnetic order parameters are expressed in terms of the quasiclassical Green's functions via the self-consistency equations
\begin{align}
\Delta &= \frac{\lambda_{\text{s}}}{16} \int_{-\Omega_{\text{s}}}^{\Omega_{\text{s}}} \mathrm{d} \varepsilon \, \big\langle \mathrm{Tr} \big\{ \check{\mathrm{X}}_{323} (\check{g}^{\text{R}} - \check{g}^{\text{A}}) \big\} \big\rangle \tanh (\varepsilon /2T) \,, \label{eqn:SelfConsDelta} \\
m &= \frac{\lambda_{\text{m}}}{16} \int_{-\Omega_{\text{m}}}^{\Omega_{\text{m}}} \mathrm{d} \varepsilon \, \big\langle \mathrm{Tr} \big\{ \check{\mathrm{X}}_{123} (\check{g}^{\text{R}} - \check{g}^{\text{A}}) \big\} \big\rangle \tanh (\varepsilon /2T) \,, \label{eqn:SelfConsM}
\end{align}
where the angle brackets mean averaging over the directions along the Fermi surface, i.e.,~${\langle (\ldots) \rangle = (2 \pi)^{-1} \int (\ldots) \, \mathrm{d} \phi}$.

We assume that the layers in the considered quasi-two-dimensional crystal lay in the $x$\nobreakdash-$z$\nobreakdash-plane, the magnetization in the SDW~$\mathbf{M}$ is oriented along the $z$\nobreakdash-axis, $\mathbf{M} = (0, 0, M)$, and all quantities depend on the $x$\nobreakdash-coordinate. The retarded and advanced quasiclassical Green's functions $\check{g}^{\text{R}(\text{A})}(\varepsilon)$ are related to the functions~$\check{g}(\omega)$ in the Matsubara representation if the frequency~$\omega$ is replaced as~${\omega \to \pm \mathrm{i} \varepsilon}$.

\section{Stationary Case}
\label{sec:GL-like_Eq}
\begin{figure}
\includegraphics[width=0.45\textwidth]{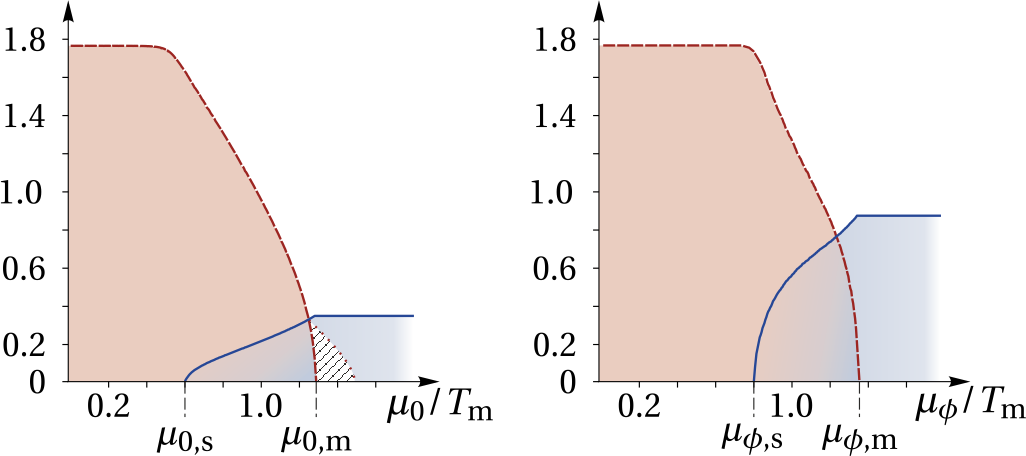}
\caption{(Color online.) Definition of~$\mu_{\text{c}}$. The dependence of~$m$ (dashed red line) respectively~$\Delta$ (solid blue line) on~$\mu_0$ (left) respectively~$\mu_{\phi}$ (right) is shown (all quantities are normalized to~$T_{\text{m}}$). In the region between~$\mu_{0,\text{s}}$ and~$\mu_{0,\text{m}}$ respectively between~$\mu_{\phi,\text{s}}$ and~$\mu_{\phi,\text{m}}$ both order parameters coexist. The $\mu_{\text{c}}$ in the text is the magnetic critical nesting parameter $\mu_{\text{c}} = \mu_{0,\text{m}} + \mu_{\phi,\text{m}} \cos (2 \phi)$. We used $T_{\text{m}}/T_{\text{s}} = 5$ and $\mu_{\phi}/T_{\text{m}} = 1.26$ on the left, and $T_{\text{m}}/T_{\text{s}} = 2$ and $\mu_{0}/T_{\text{m}} = 0.82$ on the right. The hatched region on the left denotes the inhomogeneous SDW phase, cf.~Section~\ref{sec:Discussion}.}
\label{fig:Defn_mu_critical}
\end{figure}

First, we consider the stationary equilibrium case. We derive an equation describing the coordinate dependence of the magnetic order parameter in the vicinity of the points~$\mu_{0,\text{m}}$ and~$\mu_{\phi,\text{m}}$ defined by ${m(\mu_{0,\text{m}}) = 0}$ and ${m(\mu_{\phi,\text{m}}) = 0}$, respectively (cf.~Fig.~\ref{fig:Defn_mu_critical}). Note that the values~$\mu_{0,\text{m}}$ and~$\mu_{\phi,\text{m}}$ are related with each other by the condition of coexistence of the superconducting and magnetic OPs (see Fig.~\ref{fig:mu0muPhiPlane}).

In the equilibrium case considered in this section we use the quasiclassical Green's functions in the Matsubara representation. We expand these matrix function~$\check{g}$ (see Eq.~(\ref{eqn:SolutionForQuasiclGF})) in the small parameter~$m$, using the self-consistency equations~(\ref{eqn:SelfConsDelta}) and~(\ref{eqn:SelfConsM}), and by solving Eq.~(\ref{eqn:Eilenberger_general}). In zero order approximation we have
\begin{equation}
\check{g}_{0} = \zeta_{\text{c}}^{-1} \check{\Lambda}_{\Delta, \text{c}} \,, \quad\text{with} \quad \check{\Lambda}_{\Delta,\text{c}} = \omega \check{\mathrm{X}}_{030} + \Delta_{\text{c}} \check{\mathrm{X}}_{323} \,,
\label{g0}
\end{equation}%
where ${\zeta_{\text{c}} = \sqrt{\omega^{2} + \Delta_{\text{c}}^{2}}}$ and~$\Delta_{\text{c}}$ is the superconducting~OP at the~QCP. The matrix~$\check{g}_{0}$ obeys the normalization condition ${\check{g}_{0} \cdot \check{g}_{0} = \check{1}}$ and does not depend neither on the coordinate nor on the critical value of~$\mu_{\text{c}}$ defined by ${m(\mu_{\text{c}}) = 0}$, see Fig.~\ref{fig:Defn_mu_critical}.

Expanding~${\check{g} = \check{g}_0 + \check{g}_1 +\check{g}_2+\check{g}_3 + \ldots}$ up to the third order, we obtain  from Eq.~(\ref{eqn:Eilenberger_general}) a recursive equation for the (${n+1}$)st correction for~$\check{g}$ which can be solved using Eq.~(\ref{g0}) as input, see Appendix~\ref{sec:Expansion}.

We will obtain the terms in~$\check{g}_2$ and~$\check{g}_3$, which do not contain the derivative, i.e., the term proportional to~$k$, in an easier way (see below). Thus, we can concentrate on the derivative term~$\check{g}_{3,k}$ only and, calculating the third order correction and keeping in eye the gradient term only, we obtain
\begin{equation}
\check{g}_{3,k} = v_{x}^2 \frac{\zeta_{\text{c}} (\zeta_{\text{c}}^2 - 3 \mu_{\text{c}}^2)}{4 D_{\text{c}}^3} \partial_x^2 m \cdot \check{X}_{123} \,,
\label{eqn:GradientTerm}
\end{equation}
where~${D_{\text{c}} = \zeta_{\text{c}}^2 + \mu_{\text{c}}^2}$ and ${v_{x} = v \cos(\theta)}$ with the modulus of the Fermi velocity~$v$ and the angle between the $x$\nobreakdash-axis and the orientation of the elliptic Fermi surface~$\theta$. This term will be included in the homogeneous equation for~$m$ on a later stage (see below).

Now, knowing the quasiclassical Green's function matrix Eq.~(\ref{eqn:SolutionForQuasiclGF}) in the homogeneous stationary case, we obtain the terms in the equation for~$m$ in the following way.

We rewrite the self-consistency equations~(\ref{eqn:SelfConsDelta}) and~(\ref{eqn:SelfConsM}) in the form
\begin{align}
\Delta / \lambda_{\text{s}} &= 2 \pi T \sum_{\omega = 0}^{\Omega_{\text{s}}} \langle g_{323} \rangle \label{eqn:SelfCons_RewrittenDelta} \\
\intertext{and}
m / \lambda_{\text{m}} &= 2 \pi T \sum_{\omega = 0}^{\Omega_{\text{m}}} \langle g_{123} \rangle \label{eqn:SelfCons_RewrittenM} \,,
\end{align}
where ${g_{323} = \mathrm{Tr} \big\{ \check{X}_{323} \cdot \check{g} \big\} / 8}$, ${g_{123} = \mathrm{Tr}\big\{ \check{X}_{123} \cdot \check{g} \big\} / 8}$,  and~$\Omega_{\text{s}/\text{m}}$ are the frequency cut-offs for the superconducting and the magnetic order parameters, respectively. We expand the coefficients~$g_{123}$ and~$g_{323}$ (see Eqs.~(\ref{eqn:Coeff_g123}) and~(\ref{eqn:Coeff_g323})) in the small parameter~$m^2$ and obtain
\begin{align}
g_{123} &\approx m \big[ \zeta D^{-1} + m^2 \zeta_{\text{c}} D_{\text{c}}^{-3} \big( 3 \mu_{\text{c}}^2 - \zeta_{\text{c}}^2 \big) / 2 \big] \label{eqn:g123Expanded} \\
\intertext{and}
g_{323} &\approx \Delta \big[ \zeta^{-1} + m^2 \zeta_{\text{c}}^{-1} D_{\text{c}}^{-2} \big(\mu_{\text{c}}^2 - \zeta_{\text{c}}^2 \big) / 2 \big] \label{eqn:g323Expanded} \,,
\end{align}
where~${\zeta = \sqrt{\Delta^2 + \omega^2}}$ and~${D = \zeta^2 + \mu^2}$. Equations~(\ref{eqn:SelfCons_RewrittenDelta}) and~(\ref{eqn:SelfCons_RewrittenM}) acquire the form
\begin{align}
\lambda_{\text{s}}^{-1} &= 2 \pi T \sum_{\omega = 0}^{\Omega_{\text{s}}} \big\langle \zeta^{-1} + m^2 \zeta_{\text{c}}^{-1} D_{\text{c}}^{-2} \big(\mu_{\text{c}}^2 - \zeta_{\text{c}}^2 \big) / 2 \big\rangle \label{eqn:SelfCons_ApproxDelta} \\
\intertext{and}
\lambda_{\text{m}}^{-1} &= 2 \pi T \sum_{\omega = 0}^{\Omega_{\text{m}}} \big\langle \zeta D^{-1} + m^2 \zeta_{\text{c}} D_{\text{c}}^{-3} \big( 3 \mu_{\text{c}}^2 - \zeta_{\text{c}}^2 \big) / 2 \big\rangle \label{eqn:SelfCons_ApproxM} \,.
\end{align}

\subsection{Determining the coexistence region}

\begin{figure}
\includegraphics[width=0.35\textwidth]{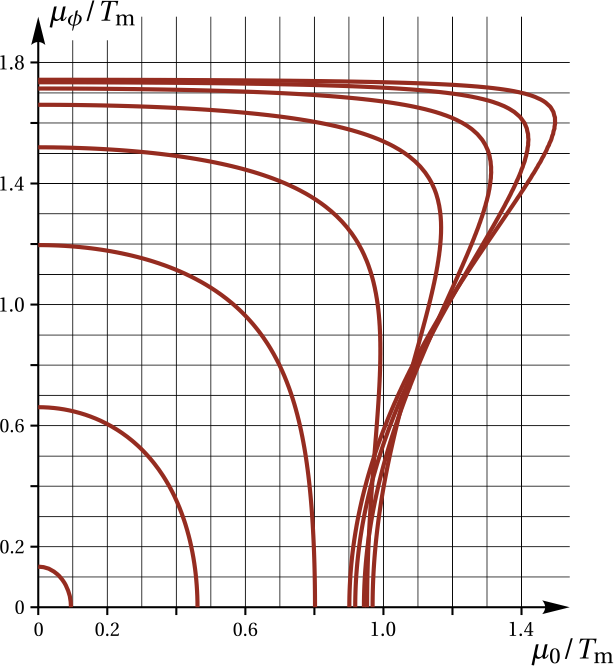}
\caption{(Color online.) Curves in the $(\mu_0, \mu_{\phi})$-plane along which ${m = 0}$ and ${\Delta = \Delta_{\text{c}}}$ for different values of ${r = T_{\text{m}} / T_{\text{s}}}$ at low temperatures. The value of~$r$ increases from bottom left to top right: ${r \approx 1.01}$, ${r = 1.1}$, ${r = 1.4}$, ${r = 2.0}$, ${r = 3.2}$, ${r = 4.9}$, ${r = 7.5}$, ${r = 10.0}$.}
\label{fig:mu0muPhiPlane}
\end{figure}

In order to find the region of existence of the QCP in terms of the nesting parameter, i.e., the region in the $(\mu_{0}, \mu_{\phi})$\nobreakdash-plane where~${m=0}$ and~${\Delta = \Delta_{\text{c}}}$, we have to set ${m = 0}$ in Eqs.~(\ref{eqn:SelfCons_ApproxDelta}) and~(\ref{eqn:SelfCons_ApproxM}). Then, the values of~$\Delta_{\text{c}}$ and~$\mu_{\text{c}}$ at the magnetic QCP, i.e., at the point where ${m = 0}$, are determined by the equations
\begin{align}
\lambda_{\text{s}}^{-1} &= 2 \pi T \sum_{\omega = 0}^{\Omega_{\text{s}}} \zeta_{\text{c}}^{-1} \label{eqn:DeltaAtMagnQCP} \\
\intertext{and}
\lambda_{\text{m}}^{-1} &= 2 \pi T \sum_{\omega = 0}^{\Omega_{\text{m}}} \big\langle \zeta_{\text{c}} D_{\text{c}}^{-1} \big\rangle \label{eqn:MAtMagnQCP} \,.
\end{align}
At low temperatures we find from Eqs.~(\ref{eqn:DeltaAtMagnQCP}) and~(\ref{eqn:MAtMagnQCP})
\begin{equation}
\big\langle s_{\text{c}} \rho_{\text{c}}^{-1} \ln (s_{\text{c}} + \rho_{\text{c}}) \big\rangle = \ln (r) \label{eqn:M_0_Curve}\,,
\end{equation}
where~${s_{\text{c}} = \mu_{\text{c}} / \Delta_{\text{c}}}$, ${\rho_{\text{c}} = \sqrt{1 + s_{\text{c}}^2}}$ and~${r = T_{\text{m}} / T_{\text{s}}}$ with the critical temperatures of the antiferromagnetic~($T_{\text{m}}$) when~${\Delta = 0}$ and the superconducting~($T_{\text{s}}$) when~${m = 0}$ transitions. This equation couples the values of the superconducting and the magnetic order parameters at zero temperature related to~$T_{\text{s}}$ and~$T_{\text{m}}$, respectively, with the critical value of the nesting parameter~${\mu_{\text{c}} = \mu_{0,\text{c}} + \mu_{\phi,\text{c}} \cos (2 \phi)}$, i.e., for a fixed ratio~${r = T_{\text{m}} / T_{\text{s}}}$ we find a curve in the $(\mu_0, \mu_{\phi})$\nobreakdash-plane along which ${m = 0}$ and ${\Delta = \Delta_{\text{c}}}$. Actually, we have to introduce a third parameter describing the ratio between the superconducting transition temperature~$T_{\text{s}}$ and the critical value of the superconducting order parameter at the magnetic QCP~$\Delta_{\text{c}}$. However, since, for~${m = 0}$, as follows directly from the self-consistency equation for~$\Delta$ (see also, e.g., Refs.~(\onlinecite{Chubukov10}) and~(\onlinecite{Schmalian10})), $T_{\text{s}}$~is independent of~$\mu$, we can take its value at~${\mu = 0}$ leading to the BCS\nobreakdash-like relation between~$\Delta_{\text{c}}$ and~$T_{\text{s}}$. In Fig.~\ref{fig:mu0muPhiPlane} we plot the curve defined by~Eq.~(\ref{eqn:M_0_Curve}) for different values of~$r$. At this point our results reproduce those obtained by Vavilov~\textit{et~al}.\cite{VavilovChubukovVorontsov10}
Remarkable is the fact that~$\mu_{\phi,\text{c}}$ becomes a multivalued function of~$\mu_{0,\text{c}}$ with increasing~$r$. Also, the allowed values of~$\mu_{0,\text{c}}$ and~$\mu_{\phi,\text{c}}$ are limited to about ${\mu_{0,\text{c}} \lesssim 1.7 T_{\text{m}}}$ and ${\mu_{\phi,\text{c}} \lesssim 1.8 T_{\text{m}}}$ for all~$r$.

\subsection{Time-independent Ginzburg--Landau-like equation}

Now, we proceed with the derivation of the stationary equation for~$m$. Subtracting Eq.~(\ref{eqn:DeltaAtMagnQCP}) from Eq.~(\ref{eqn:SelfCons_ApproxDelta}), and Eq.~(\ref{eqn:MAtMagnQCP}) from Eq.~(\ref{eqn:SelfCons_ApproxM}) we arrive at
\begin{align}
0 &= 2 \pi T \sum_{\omega = 0}^{\infty} \big\langle \zeta^{-1} - \zeta^{-1}_{\text{c}} + m^2 \zeta_{\text{c}}^{-1} D_{\text{c}}^{-2} \big(\mu_{\text{c}}^2 - \zeta_{\text{c}}^2 \big) / 2 \big\rangle \label{eqn:SubtractedEquationDelta} \\
\intertext{and}
0 &= 2 \pi T \sum_{\omega = 0}^{\infty} m \big\langle \zeta D^{-1} -\zeta_{\text{c}} D_{\text{c}}^{-1} + m^2 \zeta_{\text{c}} D_{\text{c}}^{-3} \big( 3 \mu_{\text{c}}^2 - \zeta_{\text{c}}^2 \big) / 2 \big\rangle \label{eqn:SubtractedEquationM} \,,
\end{align}
where ${\zeta_{\text{c}} = \sqrt{\omega^{2} + \Delta_{\text{c}}^{2}}}$ and the sum now runs to infinity. The terms in the sums coincide with~${\mathrm{Tr}\big\{ \check{X}_{323} \cdot (\check{g}_{1} + \check{g}_{2} + \check{g}_{3}) \big\} / 8}$ and~${\mathrm{Tr}\big\{ \check{X}_{123} \cdot (\check{g}_{1} + \check{g}_{2} + \check{g}_{3}) \big\} / 8}$, respectively, calculated following the scheme in Appendix~\ref{sec:Expansion} but without the gradient term. The gradient term in the equation describing the behavior of~$m$ in the vicinity of the magnetic QCP is obtained adding the term~${\mathrm{Tr}\big\{ \check{X}_{123} \cdot \check{g}_{3,k} \big\} / 8}$, see Eq.~(\ref{eqn:GradientTerm}), to the right hand side of Eq.~(\ref{eqn:SubtractedEquationM}), i.e.,
\begin{align}
0 = 2 \pi T \sum_{\omega = 0}^{\infty} &\big\langle \big( \zeta D^{-1} -\zeta_{\text{c}} D_{\text{c}}^{-1} \big) m \notag \\
&+ \zeta_{\text{c}} D_{\text{c}}^{-3} 2^{-1} \big( 3 \mu_{\text{c}}^2 - \zeta_{\text{c}}^2 \big) \big( m^3 - 2^{-1} v_{x}^2 \partial_{x}^2 m \big) \big\rangle \label{eqn:SubtractedEquationMWithGradient} \,.
\end{align}
This equation contains terms of higher order than the here considered third order correction and thus we expand in the next step the terms ${\zeta^{-1} - \zeta^{-1}_{\text{c}}}$ and ${\zeta D^{-1} -\zeta_{\text{c}} D_{\text{c}}^{-1}}$ in~${\delta\Delta = \Delta - \Delta_{\text{c}}}$ and~${\delta\mu = \mu - \mu_{\text{c}}}$ obtaining \begin{align}
\zeta^{-1} - \zeta^{-1}_{\text{c}} &\simeq - \zeta_{\text{c}}^{-3} \Delta_{\text{c}} \delta\Delta \\
\intertext{and}
\zeta D^{-1} -\zeta_{\text{c}} D_{\text{c}}^{-1} &\simeq - D_{\text{c}}^{-2} \big( 2 \zeta_{\text{c}} \mu_{\text{c}} \delta\mu + \zeta_{\text{c}}^{-1} \big(\zeta_{\text{c}}^2 - \mu_{\text{c}}^2\big) \Delta_{\text{c}} \delta\Delta \big) \,,
\end{align}
respectively. By virtue of these relations we can eliminate~$\delta\Delta$ from Eqs.~(\ref{eqn:SubtractedEquationDelta}) and~(\ref{eqn:SubtractedEquationMWithGradient})---it follows from the first of them that
\begin{equation}
\Delta_{\text{c}} \delta\Delta = - \kappa m^2
\label{eqn:Delta_M_Relation}
\end{equation}
with
\begin{equation}
\kappa = A_1 / (2 A_0) \,, \label{eqn:DefnKappa}
\end{equation}
where
\begin{align}
A_1 &= 2 \pi T \sum_{\omega = 0}^{\infty} \big\langle D_{\text{c}}^{-2}  \zeta_{\text{c}}^{-1} \big( \zeta_{\text{c}}^2 - \mu_{\text{c}}^2 \big) \big\rangle \\
\intertext{and}
A_0 &= 2 \pi T \sum_{\omega = 0}^{\infty} \zeta_{\text{c}}^{-3} \,.
\end{align}

\begin{figure*}
\includegraphics[width=\textwidth]{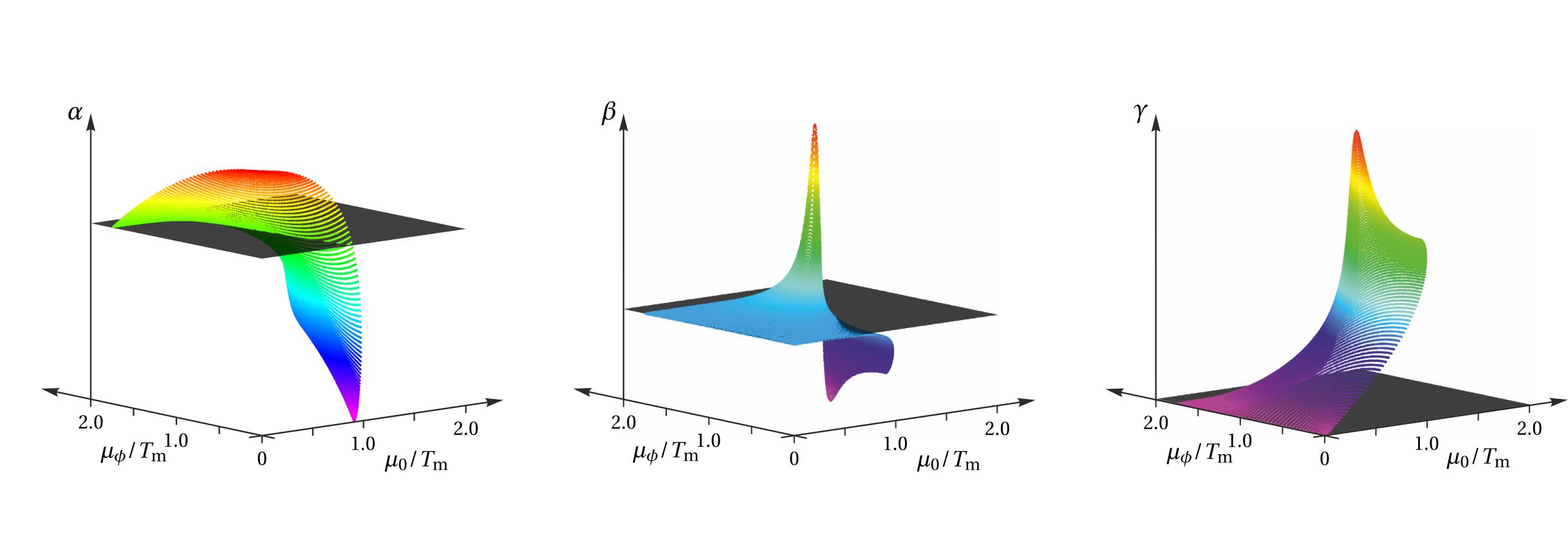}
\caption{(Color online.) The coefficients~$\alpha$, $\beta$, and~$\gamma$ in arbitrary units along the trajectories in the ($\mu_0,\mu_{\phi}$)-plane at the critical point, as calculated from Eq.~(\ref{eqn:M_0_Curve}) for different ${r = T_{\text{m}} / T_{\text{s}}}$, cf. Fig.~\ref{fig:mu0muPhiPlane}. The black surface highlights the zero level. The coefficient~$\gamma$ is always positive, whereas~$\alpha$ and~$\beta$ change their sign depending on~$\mu_0$ and~$\mu_{\phi}$ as well as on~$r$. Note that for sufficiently close transition temperatures (${r \lesssim 1.4}$) $\alpha$~is positive along the corresponding trajectory. Also, $\alpha$~depends on the angle~$\theta$---plotted is~$\alpha(\theta = 0)$ as being the most restrictive value, cf.~Fig~\ref{fig:AllowedRegion} and discussion in the text.}
\label{fig:Alpha_Beta_Gamma}
\end{figure*}

Thus, we arrive at the equation for~$m$ solely, which has the form of a Ginzburg--Landau equation,
\begin{equation}
- \alpha \partial^2_{\tilde{x}} \tilde{m} + \big( \gamma \delta\tilde{\mu} + \beta \tilde{m}^2 \big) \tilde{m} = 0 \,,
\label{eqn:TimeIndependentEquationForM}
\end{equation}
where we introduced the dimensionless quantities~${\tilde{x} = x / l}$ with~${l = v / T_{\text{m}}}$, ${\delta\tilde{\mu} = \delta\mu / T_{\text{m}}}$ and~${\tilde{m} = m / T_{\text{m}}}$, and the coefficients~$\alpha$, $\beta$ and~$\gamma$ are defined in the Appendix~\ref{sec:Coefficients} in terms of the microscopic parameters of the system.

In Fig.~\ref{fig:Alpha_Beta_Gamma} we show the dependence of these coefficients on the parameters~$\mu_{0,\phi}$ and~$r$. For a given material these parameters are determined by the doping level. The region of the coexistence of the superconducting and magnetic order parameters corresponds to those values of~$\mu_{0,\phi}$ and~$r$, for which the coefficients~$\alpha$ and~$\beta$ are positive (the coefficient~$\gamma$ is always positive). It follows from the correspondence of the coefficients~$\alpha$, $\beta$, and~$\gamma$ with the coefficients of the expansion of the free energy in the small parameter~$m$, that if ${\beta > 0}$, the quantum phase transition takes place at~${\mu = \mu_{\text{c}}}$. Negative~$\beta$ corresponds to the first order transition. In the case of positive~$\beta$ a uniform SDW state is stable provided that ${\alpha > 0}$. Otherwise the amplitude of the SDW~$m$ will be modulated in space with some wave vector~$q$ and an incommensurate SDW or the so-called soliton phase\cite{Gorkov_Teitelbaum_2010} may arise in the system (see the next section).

Although Eq.~(\ref{eqn:TimeIndependentEquationForM}) looks like the stationary Ginzburg--Landau equation or, in a more general case, like an equation which describes the spatial dependence of an order parameter in the vicinity of the critical temperature~$T_{\text{c}}$,\cite{Landau_Khalatnikov_54,L-L10} there is an essential difference between these two types of equations. In the Landau approach to the description of the phase transition of the second type near the critical temperature~$T_{\text{c}}$, the second term in Eq.~(\ref{eqn:TimeIndependentEquationForM}) contains a small parameter~${(T_{\text{c}}-T )}$ and the temperature~$T$ is assumed to be close to~$T_{\text{c}}$, whereas in the case under consideration the deviation from the critical point (critical doping),~$\delta \mu $, is a small parameter and~$T$ can be much smaller than $T_{\text{c}}$.

The key point in the derivation of Eq.~(\ref{eqn:TimeIndependentEquationForM}) is the presence of the energy scale~${\Delta_{\text{c}} \gg m}$ in the coexistence region, due to which an expansion of the Green's functions in the small parameter~$m/\Delta_{\text{c}}$ is possible and, thus, the derived equation is valid at arbitrary temperatures~${T < T_{\text{s}} < T_{\text{m}}}$.

Equation~(\ref{eqn:Delta_M_Relation}) shows that the spatial (and temporal) behavior of the superconducting order parameter~$\Delta$ is determined by the dependence~$m(x)$ (or~$m(x,t)$ in the non-stationary case) of the SDW order parameter.

Note that in Eq.~(\ref{eqn:TimeIndependentEquationForM}) we assume~${\delta\mu \equiv \delta\mu_0}$---otherwise one should include~${\delta\mu = \delta\mu_0 + \delta\mu_{\phi} \cos( 2 \phi)}$ into the averaging procedure, i.e., the first term in the brackets has in general the form (cf.~Eq.~(\ref{eqn:GammaTemperature}))
\begin{equation}
2 \pi T \sum_{\omega = 0}^{\infty} \big\langle 2 D_{\text{c}}^{-2} \mu_{\text{c}} \zeta_{\text{c}} \delta\mu \big\rangle \,.
\end{equation}

In Section~\ref{sec:TDGL} we derive the time-dependent equation for~$m$ and investigate the stability of a uniform SDW state.

\section{\label{sec:TDGL}Time-dependent Equation for the magnetic order parameter near the magnetic QCP}
\label{sec:GL-like_Eq_timedependent}

In contrast to ordinary superconductors with two characteristic energies,~$\Delta$ and~$T$, in the considered case of superconductors with an SDW there exist three characteristic parameters, i.e.,~$m$,~$\Delta$ and~$T$. As is well known, the stationary Ginzburg--Landau equations for~$\Delta$ in ordinary superconductors can be obtained by using an expansion of the free energy in powers of a small parameter~$\Delta/T$ near the critical temperature~$T_{\text{c}}$, i.e., in the region ${|T - T_{\text{c}} | \ll T_{\text{c}}}$.\cite{GL,Gorkov_59a,*Gorkov_59b} This expansion was employed by Landau who developed a rather general theory of the phase transition of the second type.\cite{Landau_37}

Generalization of the Ginzburg--Landau equations to a non-stationary case is not an easy task. Moreover, in a general case one can not obtain a closed equation for the time-dependent~$\Delta(t)$. The point is that near~$T_{\text{c}}$ the order parameter~$\Delta $ is determined by the quasiparticle distribution function~$n(\varepsilon,t)$ the
relaxation of which is due to inelastic scattering processes. Therefore, one needs to solve a complicated kinetic equation for~$n(\varepsilon,t)$ with account for inelastic processes. In a general case it is not possible to find a solution for the distribution function in terms of the time-dependent~$\Delta(t)$.

Using the technique of thermal Green's functions and making a subsequent analytical continuation to real frequencies, Gor'kov and Eliashberg have shown that only in an exotic case of superconductors with a large concentration of paramagnetic impurities the time-dependent Ginzburg--Landau equation can be obtained.\cite{G-Eliash} Namely, the generalization of the Ginzburg--Landau equation to a non-stationary case is possible only in the extreme case of a high density of paramagnetic impurities (${T_{\text{c}} \tau_{\text{sf}} \ll 1}$, where~$\tau_{\text{sf}}$ is a spin-flip scattering time) when the gap in the excitation spectrum disappears (the so-called gapless superconductivity).

In the considered case the expansion is performed with respect to another small parameter, i.e.,~$m(t) / \Delta$, whereas the temperature~$T$ is assumed to be low, ${T \ll \Delta}$. Then, the number of quasiparticles is small and their influence on the amplitudes of the SDW~$m$ is negligible.

Note an important point. The dynamic equation for~$m(t)$ is analogous to Eq.~\ref{eqn:Eta}, but not to the non-stationary Ginzburg--Landau equation. The latter describes the dynamics of a complex OP~$\Delta$, i.e., the temporal evolution of the modulus~$|\Delta|$ and phase~$\chi$ of the OP~$\Delta$ or, to be more exact, it describes the behavior of~$|\Delta|$ and of the gauge-invariant quantity~${\mu_{\chi} = \partial_t \chi + 2 e V}$, where~$V$ is the electrical potential. Besides, the expression for the current contains a gauge-invariant velocity of the condensate. In contrast to ordinary superconductors, in the case under consideration~$m$ is a real quantity.

The time-dependent equation for~$m$ can be used to investigate the time evolution of the order parameters involved\cite{Kopnin} (e.g., in alternating external fields, in treating transport phenomena or also problems with dissipative terms induced by a d.c.~electric field) as well as to analyze the stability of the stationary solutions.

Unlike the Gor'kov's and Eliashberg's approach based on the use of the thermal Green's functions,\cite{G-Eliash} we will apply the method of time-dependent Green's functions (in other words, the real-time Green's functions) in the Keldysh technique.\cite{Keldysh_1965} In this method, there is no need to use the analytical continuation. We rewrite the self-consistency equation~(\ref{eqn:SelfConsM}) in the form
\begin{equation}
m(t) = \lambda_{\text{m}} \int \mathrm{d} t_1 \, \big\langle g_{123}^{\text{R}} (t, t_1) F(t_1 - t) - F(t - t_1) g_{123}^{\text{A}} (t_1, t) \big\rangle \,,
\label{eqn:SelfConsMOnTime}
\end{equation}
where~$F(t)$ is the equilibrium distribution function with the Fourier transform ${F(\varepsilon) = \tanh(\varepsilon / 2T) = \int \mathrm{d} t \ F(t) \exp(\mathrm{i} \varepsilon t)}$ and~$g_{123}^{\text{R,A}}$ are the retarded~(advanced) quasiclassical Green's functions which can be obtained from the thermal Green's functions used in Section~\ref{sec:GL-like_Eq} in equilibrium case if we make a substitution ${\omega \rightarrow -\mathrm{i} (\varepsilon \pm \mathrm{i}0 )}$ for~$g_{123}^{\text{R(A)}}$. Further, we take into account that actual frequencies~${\Omega \sim (kv)^2 \sim \delta\mu}$ (where~$k$ is a characteristic wave vector) near the QCP are much lesser than the characteristic energies in the integral of Eq.~(\ref{eqn:SelfConsMOnTime}), ${\varepsilon_{\text{ch}} \sim \mathrm{max} \{ T, \Delta \}}$. Thus, going over to the Fourier components one can write Eq.~(\ref{eqn:SelfConsMOnTime}) as follows
\begin{align}
m(t) &= 2 \pi \lambda_{\text{m}} \int \mathrm{d} \varepsilon \, \big\langle \big[ g_{123}^{\text{R}} (\varepsilon, t) - g_{123}^{\text{A}}(\varepsilon, t) \big] \notag \\
&- 2^{-1} \mathrm{i} \partial_t \big[ g_{123}^{\text{R}}(\varepsilon, t) + g_{123}^{\text{A}}(\varepsilon, t) \big] \, \partial_\varepsilon \tanh(\varepsilon/2T) \big\rangle \,.
\label{eqn:SelfConsMOnTime1}
\end{align}
Integrating the second term by parts we obtain in the Matsubara representation
\begin{equation}
m(t) / \lambda_{\text{m}} = 2 \pi T \sum_{0}^{\Omega_{m}} \big\langle g_{123}(\omega, t) + 2^{-1} \partial_t m \, \partial_\omega \zeta_{\text{c}} D_{\text{c}}^{-1} \big\rangle \,,
\label{eqn:SelfConsMOnTime2}
\end{equation}
where~$g_{123}(\omega, t)$ is the stationary part found above, see Eq.~(\ref{eqn:Coeff_g123}) and Eq.~(\ref{eqn:GradientTerm}). Now, we have to take into account that~$m$ slowly depends on time. It can be shown that the dependence of the functions~$g_{123}(\varepsilon, t)^{\text{R,A}}$ on~$\Omega$ is much weaker than that given by the last term in Eq.~(\ref{eqn:SelfConsMOnTime2}). The reason is that there appear small corrections of higher orders, i.e., proportional to~$(\Omega/\Delta_{\text{c}})^2$.

Calculating the last term in Eq.~(\ref{eqn:SelfConsMOnTime2}) at low temperatures~($T \ll \Delta_{\text{c}}$) and adding it to the right hand side of Eq.~(\ref{eqn:TimeIndependentEquationForM}), we obtain the generalized equation, which describes the time and the space evolution of the magnetic order parameter~$m$ near the~QCP
\begin{equation}
- \alpha \partial^2_{\tilde{x}} \tilde{m} + \big( \gamma \delta\tilde{\mu} + \beta \tilde{m}^2 \big) \tilde{m} = - \varsigma \partial_{\tilde{t}} \tilde{m}
\label{eqn:TimeDependentEquationForM}
\end{equation}
with ${\varsigma = 2^{-1} T_{\text{m}} \Delta_{\text{c}}^{-1} \big\langle \big( 1 + s_{\text{c}}^2 \big)^{-1} \big\rangle}$ and~${\tilde{t} = T_{\text{m}} t}$ in the limit of low temperatures (cf.~Eq.~(\ref{eqn:VarSigmaLowTemperatures})).

As concerns the corrections of~$\delta\Delta$, the relation between~$\delta\Delta$ and~$m$ remains unchanged in the main approximation since the time derivative~$\partial_{t} \delta\Delta$ is small. The generalized relation between~$\delta\Delta$ and~$m$ has the form ${c_1 \partial_{\tilde{t}} \delta\tilde{\Delta} + \delta\tilde{\Delta} = - \kappa \tilde{m}^2}$, where~${c_1 \simeq 1}$. As follows from Eq.~(\ref{eqn:TimeDependentEquationForM}), ${|\partial_{\tilde{t}} \delta\tilde{\Delta}| \propto |\delta\tilde{\mu}\delta\tilde{\Delta}| \ll |\kappa \delta\tilde{\Delta}|}$. Thus, the time derivative of~$\delta\Delta$ can be neglected in Eq.~(\ref{eqn:TimeDependentEquationForM}).

The obtained time-dependent equation~(\ref{eqn:TimeDependentEquationForM}) looks like Eq.~(\ref{eqn:Eta}) derived by Landau and Khalatnikov\cite{Landau_Khalatnikov_54,L-L10} to describe the relaxation of an order parameter to its equilibrium value in the vicinity of the critical temperature for a second order phase transition. It allows to draw conclusions about the states of coexistence of superconducting and magnetic order parameters as well as about the stability of these phases.

Our approach is complementary to the one based on the analysis of the free energy used in Refs.~\onlinecite{Chubukov10,Schmalian10}. Indeed, we can study the stability of the states with ${m_0 = 0}$ and ${m_1 = \sqrt{- \gamma \beta^{-1} \delta\mu} = \sqrt{\gamma \beta^{-1} |\delta\mu|}}$ linearizing Eq.~(\ref{eqn:TimeDependentEquationForM}) with respect to the deviation from equilibrium~$\delta m$ written in the form ${\delta m_{0,1}(x,t) \propto \exp(\mathrm{i} \Omega t - \mathrm{i} k x)}$, where~$\delta m_{0,1}$ are the amplitudes of the deviations from these two possible states~$m_0$ and~$m_1$, respectively. Since the coefficient~${\gamma > 0}$ (see Eq.~(\ref{eqn:TimeIndependentEquationForM}) and Fig.~\ref{fig:Alpha_Beta_Gamma}) is always positive, the coefficient~$\beta$ should be positive---otherwise we would have a transition of the first order to a state with a finite amplitude---and thus the quantity~$\delta\mu$ must be negative. In the first case we obtain ${\mathrm{i} \Omega = - \gamma \varsigma^{-1} \delta\mu - \alpha \varsigma^{-1} k^2}$. The first term here is positive (${\delta\mu < 0}$), and this means that this state is unstable provided that ${\alpha > 0}$.

For the second case we find ${\mathrm{i} \Omega = 2 \gamma \varsigma^{-1} \delta\mu - \alpha \varsigma^{-1} k^2}$, i.e., this state is stable if~$\alpha > 0$. If the coefficient~$\alpha$ is negative for some~$\mu_{\text{c}}$, nonuniform perturbations will grow in time with increasing~$k$. As a result, a modulation of the SDW~amplitude would be established in the system.

It is important to note that the coefficient~$\alpha$ depends on the angle~$\theta$. For sufficiently close transition temperatures (${r \lesssim 1.4}$) this dependence is negligible, since ${\alpha > 0}$ for all relevant values of~$\mu_0$ and~$\mu_{\phi}$ independent on~$\theta$. For larger ratios of~$T_{\text{m}} / T_{\text{s}}$ there exists a segment of the critical coexistence curve where for some fixed values of parameters~$\mu_{0}$ and~$\mu_{\phi}$ lying in the region where~${\beta > 0}$ the coefficient~$\alpha$ is positive for~${\theta = \pi /2}$, but turns negative for ${\theta = 0}$ i.e., a modulation of the SDW may occur in the direction parallel to the major axis of the elliptic Fermi surface. In Fig.~\ref{fig:AllowedRegion} we indicate the dependence of~$\alpha(\theta)$ showing the ${\alpha = 0}$ line for two limiting values of~$\theta$, i.e.,~for ${\theta = 0}$ and ${\theta = \pi / 2}$.

\section{Domain wall}
\label{sec:DomainWall}

As already noted above, domain walls in real samples are related most likely not with the change of the value of~$m$, but with the rotation of the vector~$\mathbf{m}$. Nevertheless, in order to understand qualitatively the behavior of different quantities, we present a solution for the domain wall~(DW) caused by variation in space of the magnitude of the magnetic order parameter~$m$. For example, we show that the superconducting order parameter gets enhanced in the center of the~DW.

Since the coefficient~$\gamma$ is positive (cf.~Fig.~\ref{fig:Alpha_Beta_Gamma}), the deviation from the perfect nesting~$\delta \mu$ must be negative provided the coefficient~$\beta$ is positive and vice versa. The solution of this equation for a DW has the well known form\cite{L-L8,Gorkov_Teitelbaum_2010}
\begin{equation}
\tilde{m}(x) = \tilde{m}_0 \tanh( \tilde{x} / l_0) \,,
\label{eqn:DomainWall}
\end{equation}
with the amplitude of the SDW DW~${\tilde{m}_0 = \sqrt{- \gamma \beta^{-1} \delta\tilde{\mu}}}$. The width of the DW~$l_0$ equals ${l_0 = l \sqrt{- 2 \alpha \gamma^{-1} \delta\tilde{\mu}^{-1}}}$, i.e., it becomes infinitely large at the magnetic~QCP.

Note that unlike Ref.~\onlinecite{Gorkov_Teitelbaum_2010}, where the DW was studied for a spatial distribution of a scalar order parameter related to the nesting between different bands in a two-band material in the absence of superconductivity, we consider the DW in the SDW in the presence of superconductivity. Thus, the characteristic size of the DW~$l_{0}$ depends on~$\Delta$ (see Eqs.~(\ref{eqn:AlphaTemperature}) and~(\ref{eqn:GammaTemperature}) for~$\alpha$ and~$\gamma$).

\begin{figure}
\includegraphics[width=0.35\textwidth]{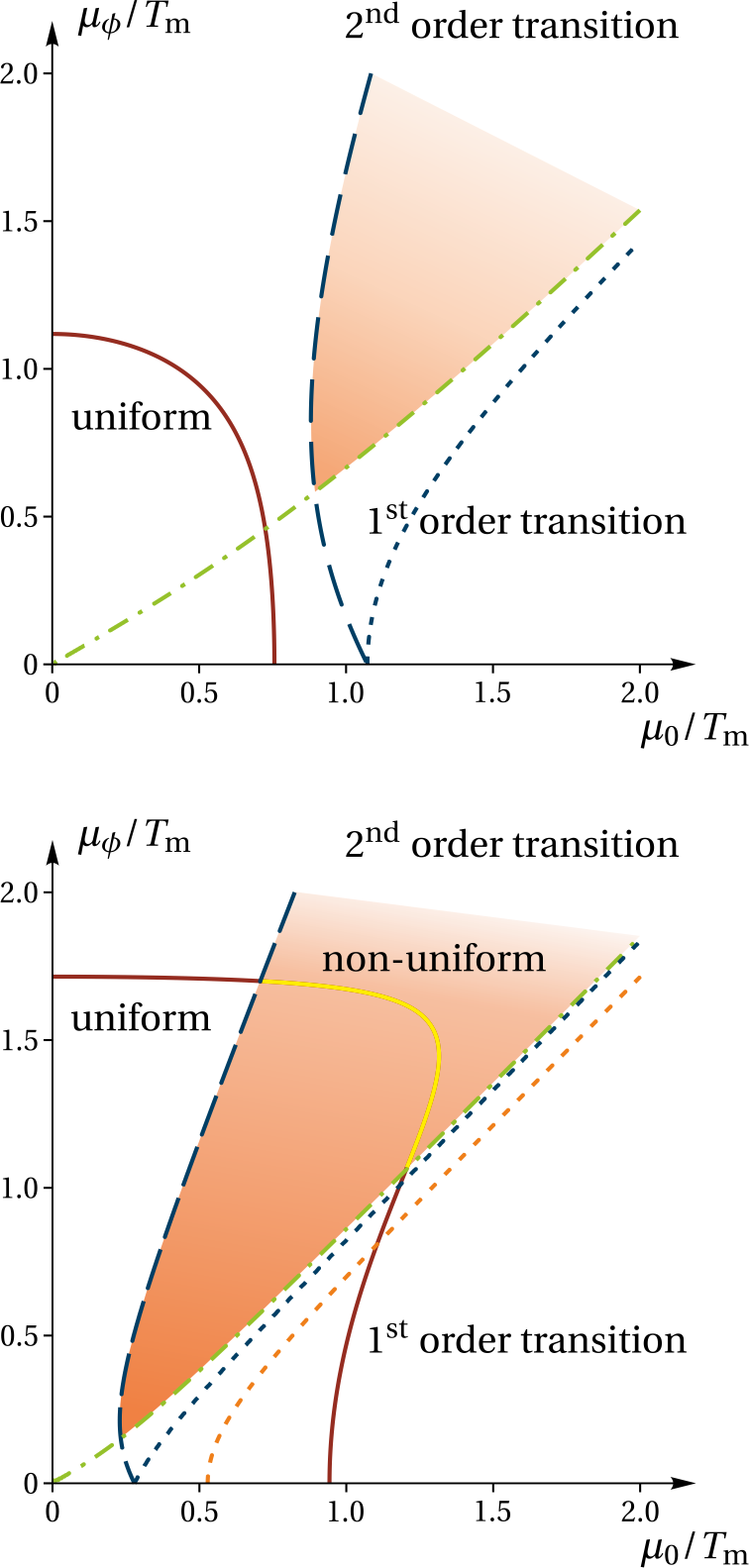}
\caption{(Color online.) Different regions in the $(\mu_0, \mu_{\phi})$-plane. We used~${r \approx 1.3}$ in the top panel and~${r \approx 10.0}$ on the bottom panel. The solid red line is one of the critical coexistence curves from Fig.~\ref{fig:mu0muPhiPlane}. The dash-dotted green line, along which ${\beta = 0}$, divides the curve into segments of first~(${\beta < 0}$) respectively second~(${\beta > 0}$) order phase transition between the SC and the SDW states. The long dashed blue line, along which ${\alpha = 0}$ (for ${\theta = 0}$), subdivides the latter into segments of uniform~(${\alpha > 0}$ independent on~$\theta$) and non-uniform~($\alpha$ changes sign as function of~$\theta$) SDW states. The dependence on~$\theta$ is indicated by the short-dashed blue line, along which ${\alpha = 0}$ now for ${\theta = \pi / 2}$ and the lines for other values of~$\theta$ lie between these two blue curves. In the uniform state the solution of Eq.~(\ref{eqn:TimeDependentEquationForM}) describes a domain wall. Note that in the case of sufficiently small~$r$ (approximately ${r < 1.4}$) $\alpha$~is always positive along the coexistence line (top panel), thus only the uniform SDW exists. In the bottom panel we also show the curve along which~${\kappa = 0}$, being positive above this line (short-dashed orange), which means a negative correction to~$\Delta_{\text{c}}$ outside the center of the domain wall.}
\label{fig:AllowedRegion}
\end{figure}

The coefficient~$\alpha$ also has to be positive in order for the solution to be valid. These two conditions restrict us to a segment on one of the lines in Fig.~\ref{fig:mu0muPhiPlane}, see Fig.~\ref{fig:AllowedRegion}.
Note that the width of this segment is, for a given~${r > 1.4}$, strongly dependent on the angle~$\theta$ as shown in Fig.~\ref{fig:AllowedRegion}, i.e., the line~${\alpha = 0}$ approaches the line~${\beta = 0}$ already if the deviation of~$\theta$ from zero is very small, thus extending the segment where the DW~solution is valid. Note also that in the vicinity of the ${\beta = 0}$ line our approach breaks down, for in this region~$m$ is very large being proportional to~$\sqrt{\beta^{-1}}$. The plots in Fig.~\ref{fig:AllowedRegion} should be read as follows. First, we determine the region where~$\alpha$ and~$\beta$ are positive respectively negative regardless the true values of~$\mu_{0,\text{c}}$ and~$\mu_{\phi,\text{c}}$ where they should be evaluated. Then, we check if the $\mu_{\text{c}}$\nobreakdash-line (calculated from Eq.~(\ref{eqn:M_0_Curve})) lies in the obtained region. Thus, we obtain a segment of the $\mu_{\text{c}}$\nobreakdash-line where the DW~solution is valid, as the values of the coefficients~$\alpha$, $\beta$ and~$\gamma$ have their meaning, strictly speaking, only along the respective critical coexistence curve.

\begin{figure}
\includegraphics[width=0.4\textwidth]{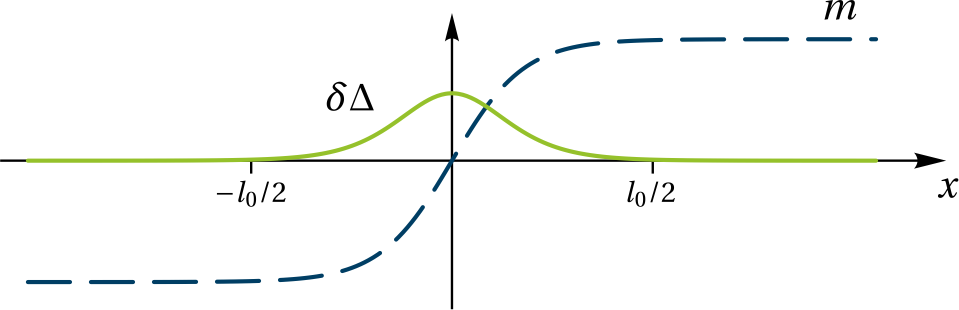}
\caption{(Color online.) The dashed blue line shows the domain wall as described by Eq.~(\ref{eqn:TimeIndependentEquationForM}). It has the width~${l_0 = l \sqrt{- 2 \alpha \gamma^{-1} \delta\tilde{\mu}^{-1}}}$. The solid green line shows the corresponding dependence~$\delta\Delta(x)$ counted from its bulk value.}
\label{fig:PlotDW}
\end{figure}

\begin{figure}
\includegraphics[width=0.4\textwidth]{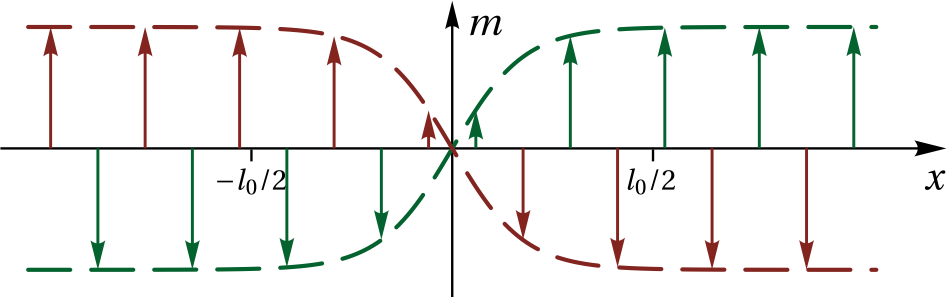}
\caption{(Color online.) The dashed line shows the domain wall in the corresponding sublattice, distinguished by the color. The behavior of the magnetization showed on Fig.~\ref{fig:PlotDW} corresponds to the green line here.}
\label{fig:AF_on_x}
\end{figure}

In Fig.~\ref{fig:PlotDW} we show an example of a DW and the corresponding coordinate dependence of~${\delta\tilde{\Delta} = - \tilde{\kappa} \tilde{m}^2}$ with~${\tilde{\kappa} = \Delta_{\text{c}}^{-1} T_{\text{m}} \kappa }$ and sketch, based on this, how the antiferromagnetic order might look like in the corresponding sublattice in Fig.~\ref{fig:AF_on_x}.

One can see that the superconducting OP~$\Delta$ has a small peak in the center of the~DW, at~${x = 0}$, where the magnetic OP turns to zero. A similar effect of enhancement of superconductivity at the ferromagnetic DW takes place also in S/F~structures.\cite{Khlyustikov_Buzdin_87,Burmistrov05} This effect is quite intelligible. Always, when we have two competing OPs, the suppression of one OP leads to an enhancement of the other OP regardless of the cause of the suppression. For example, impurity scattering suppresses the amplitude of the SDW\cite{FernandesVavilovChubukov85} and an enhancement of the critical temperature of the superconducting transition temperature~$T_{\text{s}}$ takes place. A similar effect of increasing~$T_{\text{s}}$ due to impurities was predicted long ago for quasi-one-dimensional conductors where transitions into the superconducting state and into a state with a charge density wave are possible.\cite{Efetov1981,*Efetov1981rus}

\section{Discussion}
\label{sec:Discussion}

We derived a time-dependent equation~(\ref{eqn:TimeDependentEquationForM}) which describes the relaxation and the spatial behavior of the amplitude~$m$ of the SDW in the vicinity of the magnetic~QCP, where the phase transition of the second order takes place. At this point, the amplitude~$m$ turns to zero while the superconducting OP~$\Delta_{\text{c}}$ remains finite. The derived time-dependent equation is valid at low temperatures when the number of quasiparticles is low and the quasiparticles do not affect essentially the dynamics of~$m$ and~$\Delta$. Therefore, the conditions ${\{T,m\} \ll \Delta}$ should be satisfied. However, in the stationary case the Ginzburg--Landau-like equation~(\ref{eqn:TimeIndependentEquationForM}) is valid at arbitrary temperatures. Note that we considered only the case of a fixed orientation
of the magnetization in the SDW (no rotation of the~$\mathbf{m}$ vector).

The coefficients in this equation are expressed in terms of microscopic characteristics of the system, i.e., the nesting parameter ${\mu = \mu_{0} + \mu_{\phi} \cos(2 \phi)}$, the superconducting OP at the QCP~$\Delta_{\text{c}}$, and the Fermi velocity~$v$. We found regions of the second order transition in the $(\mu_{0},\mu_{\phi})$-plane. These regions are in agreement with the results of Vorontsov~\textit{et~al}.\cite{VavilovChubukovVorontsov10} who analyzed the free energy in the system.

The derived equation is valid if the influence of fluctuations can be neglected. This requirement imposes a restriction on temperatures
\begin{equation}
\delta \mu \ll T \ll \delta \mu \left( \frac{E_{F}} {T_{\text{m}}} \right)^{2} \,,
\end{equation}
The left inequality means that quantum fluctuations can be neglected\cite{L-L10} and the right one is similar to the Ginzburg--Levanjuk criterium for the applicability of the Ginzburg--Landau equation for superconductors.\cite{L-L5,Larkin_Varlamov_2005,Sachdev}

On the basis of Eq.~(\ref{eqn:TimeIndependentEquationForM}) a solution that describes a domain wall in the stationary case has been obtained. The width of the DW diverges when ${\mu \rightarrow \mu_{\text{c}}}$ defined by the criticality line Eq.~(\ref{eqn:M_0_Curve}), see Fig.~\ref{fig:mu0muPhiPlane}. The superconducting OP increases in the center of the DW as compared to its value far away from the DW, i.e., we find an enhancement of superconductivity.

Using the time-dependent Eq.~(\ref{eqn:TimeDependentEquationForM}) we studied also the stability of the state with a uniform SDW in the vicinity of the magnetic QCP where the magnetic order parameter vanishes. It was shown that a homogeneous commensurate SDW is stable on the upper part of the curve defined by the Eq.~(\ref{eqn:M_0_Curve}) which describes the region of coexistence of superconducting and magnetic OP (see Fig.~\ref{fig:AllowedRegion}). Below this region the homogeneous SDW is unstable against perturbations with a finite wave vector~$k$ propagating along the $x$\nobreakdash-axis chosen to be the direction parallel to the major axis of the elliptic Fermi surface. This instability leads to the appearance of incommensurate~SDW\cite{VavilovChubukovVorontsov10} or of a inhomogeneous SDW similar to a
soliton lattice discussed by Gor'kov and Teitelbaum (see Ref.~\onlinecite{Gorkov_Teitelbaum_2010} and references therein).

In Fig.~\ref{fig:Defn_mu_critical} we show the region of a homogeneous stable SDW coexisting with superconductivity (${\delta\mu < 0}$) and the region of an inhomogeneous SDW (${\delta\mu > 0}$) state (compare with Fig.~1 in Ref.~\onlinecite{Gorkov_Teitelbaum_2010}). The issue of finding the form of the inhomogeneous SDW deserves a separate consideration.

\acknowledgments

We are grateful to I.~Eremin for fruitful discussions and appreciate the financial support from the DFG by the Project~EF~11/8\nobreakdash-1; A.~F.~V.~thanks the DIPC for hospitality and financial support during his stay at the~DIPC.

\appendix

\section{Quasiclassical Green's functions---full expressions}
\label{app:QCGF}

\begin{figure}
\includegraphics[width=0.3\textwidth]{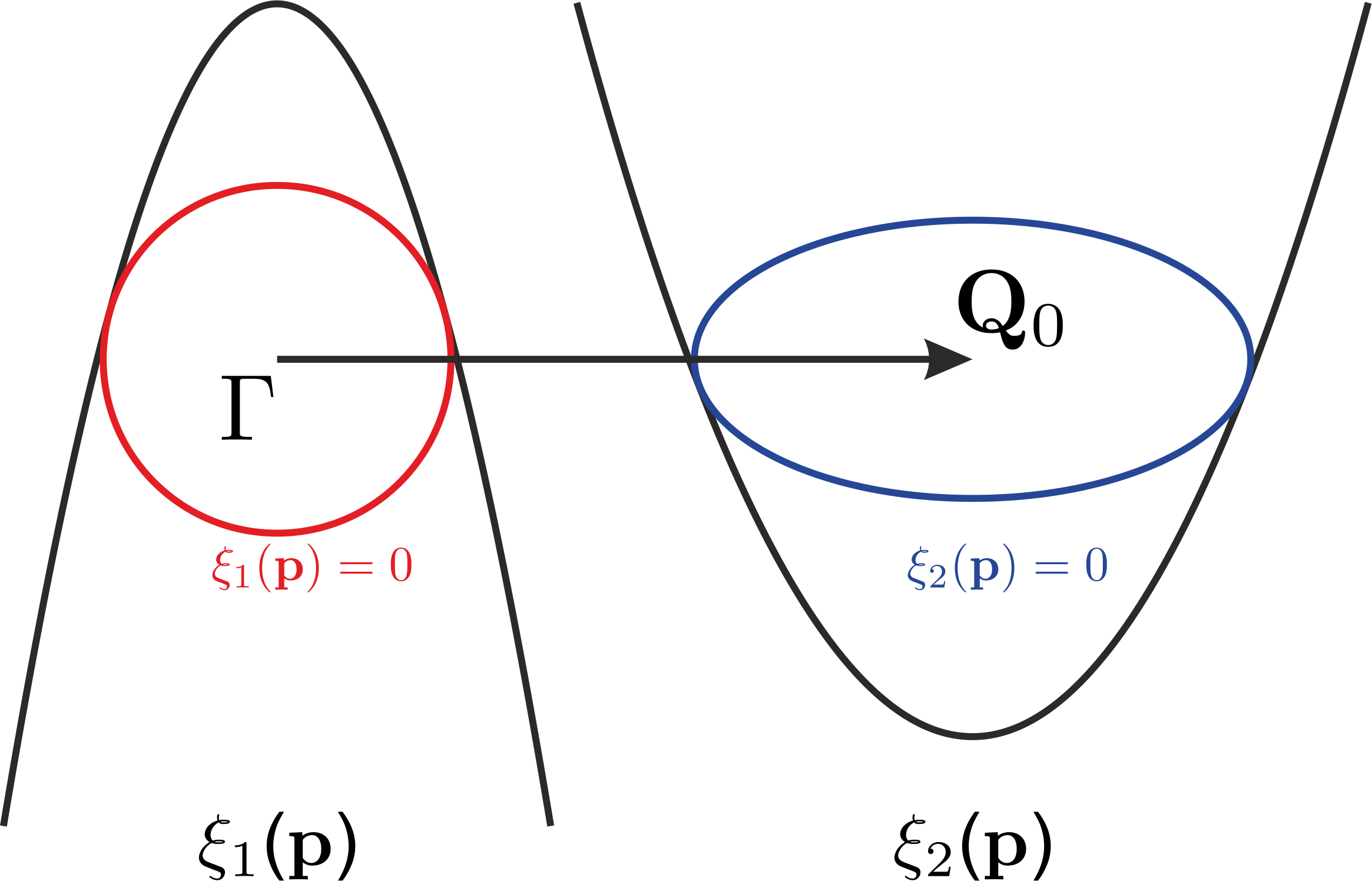}
\caption{(Color online.) Considered band structure. }
\label{fig:TwoBands}
\end{figure}

The dispersions in the hole~(1) and electron~(2) bands have the form
\begin{equation}
\label{eqn:Dispersions}
\xi_{1}(\mathbf{p}) = \mu_1 - \frac{\mathbf{p}^2}{2 m_1} \,, \qquad \xi_{2}(\mathbf{p}) = - \mu_2 + \frac{p_x^2}{2 m_{2,x}} + \frac{p_y^2}{2 m_{2,y}} \,,
\end{equation}
where the momenta~$\mathbf{p}$ in~$\xi_{1}$ are measured from the $\Gamma$\nobreakdash-point and
in $\xi_{2}$---from the nesting vector~$\mathbf{Q}_0$ see~Fig.~\ref{fig:TwoBands}.

Defining
\begin{align}
    \xi(\mathbf{p}) &= \frac{\xi_{2}(\mathbf{p}) - \xi_{1}(\mathbf{p})}{2} \\
\intertext{and}
    \mu &= \frac{\xi_{2}(\mathbf{p}) + \xi_{1}(\mathbf{p})}{2}
\end{align}
we can write the dispersions in the form
\begin{equation}
    \xi_{1,2}(\mathbf{p}) = \mp  \xi(\mathbf{p}) + \mu \,.
\end{equation}

The parameter~$\mu$ describing the deviation from the perfect nesting can be written as\cite{Chubukov10} ${\mu = \mu_{0} + \mu_{\phi} \cos(2\phi)}$, where~$\mu_0$ takes into account the difference in the electron and hole masses, whereas~$\mu_{\phi}$ describes the ellipticity of the electron pocket~(${m_{2,x} \neq m_{2,y}}$), they correspond up to a constant factor with the nesting parameters in the Vorontsov~\emph{et~al}.\ notations.\cite{Chubukov09,*Chubukov10} Within the quasiclassical approximation, the ellipticity can be neglected in the expression for the Fermi velocity~$\mathbf{v}$.

In the stationary and uniform case the matrix~${\check{g} = (\mathrm{i} / \pi) \int \mathrm{d} \xi \, (\check{\tau}_{3} \cdot \check{G})}$ can be found from the exact Green's function~$\check{G}$, which obeys Eq.~(3.3) in Ref.~\onlinecite{Moor11}, i.e., ${\check{G}^{-1}\cdot \check{G} = \check{1}}$, by inverting the matrix~$\check{G}^{-1}$ and integrating over~$\xi$.

In the case of the assumed $s_{+-}$\nobreakdash-pairing,\cite{Moor11} this
procedure yields the following expression for~$\check{g}$ in the Matsubara representation
\begin{equation}
     \check{g} = \sum_{mn\alpha} g_{mn\alpha} \cdot \check{X}_{m n \alpha} \,,
\label{eqn:SolutionForQuasiclGF}
\end{equation}
where only the six following coefficients~$g_{mn\alpha}$ are not zero and all the other vanish
\begin{align}
    g_{030} &= \omega \zeta^{-1} |\chi_{+}|^{-2} \left( \zeta \Re\{\chi_{+}\} + \mu \Im\{ \chi_{+} \} \right) \,, \label{eqn:Coeff_g030} \\
    g_{100} &= \mathrm{i} \Delta m \zeta^{-1} |\chi_{+}|^{-2} \Im\{ \chi_{+} \} \,, \label{eqn:Coeff_g100} \\
    g_{123} &= m |\chi_{+}|^{-2} \Re\{\chi_{+}\} \,, \label{eqn:Coeff_g123} \\
    g_{213} &= \mathrm{i} \omega m \zeta^{-1} |\chi_{+}|^{-2} \Im\{\chi_{+}\} \,, \label{eqn:Coeff_g213} \\
    g_{300} &= \mathrm{i} |\chi_{+}|^{-2} \left(
                \zeta \Im\{\chi_{+} \} - \mu \Re\{\chi_{+}\} \right) \,, \label{eqn:Coeff_g300} \\
    g_{323} &= \Delta \zeta^{-1} |\chi_{+}|^{-2} \left(
                \zeta \Re\{\chi_{+}\} + \mu \Im\{\chi_{+}\} \right) \,, \label{eqn:Coeff_g323}
\end{align}
where we defined ${\zeta = \sqrt{\Delta^2 + \omega^2}}$ and ${\chi_{+} = \sqrt{m^2 + \left( \zeta + \mathrm{i} \mu\right)^2}}$, with ${\Delta=|\Delta|}$.

Direct calculations show that the matrix~$\check{g}$ satisfies the orthogonality condition~Eq.~(\ref{eqn:NormCondition}).

%\bigskip

\section{Derivation of the third-order correction to~\texorpdfstring{$\mathbf{\check{g}}$}{g}}
\label{sec:Expansion}

Expanding~${\check{g} = \check{g}_0 + \sum_{n=1} \check{g}_{n}}$, we obtain for the (${n+1}$)st correction from Eq.~(\ref{eqn:Eilenberger_general}) in Matsubara representation ($[ \cdot \,, \cdot ]$~denoting the commutator)
\begin{equation}
\big[ \Pi \,, \check{g}_{n+1} \big] = \big[ \delta \Pi \,, \check{g}_n \big] + r_n \check{g}_n \,,
\label{eqn:g_n+1-correction}
\end{equation}
where ${\Pi = \zeta_{\text{c}} \check{g}_0 - \mathrm{i} \mu_{\text{c}} \check{X}_{300}}$, ${r_0 = 0}$ and ${r_n = \mathrm{i} v_{x} k \check{X}_{000}}$ for~${n \neq 0}$ with the projection of~$\mathbf{v}$ on the $x$\nobreakdash-axis---$v_{x}$, and ${\delta \Pi = m \check{X}_{123} - \delta\Delta \check{X}_{323} - \mathrm{i} \delta\mu \check{X}_{300}}$ with ${\delta\Delta = \Delta - \Delta_{\text{c}}}$ and ${\delta\mu = \mu - \mu_{\text{c}}}$ describing deviations from the critical values of the superconducting order parameter and nesting, respectively. From the normalization condition for~$\check{g}$, Eq.~(\ref{eqn:NormCondition}), we obtain the following relation for~${n\neq0}$ ($\{ \cdot \,, \cdot \}$ denoting the anticommutator)
\begin{equation}
0 = \sum_{i=0}^{n} \big\{ \check{g}_i \,, \check{g}_{n-i} \big\} \,.
\end{equation}
Thus, we arrive at the following recursive equation for~$\check{g}_{n+1}$
\begin{equation}
2 \zeta_{\text{c}} \check{g}_0 \check{g}_{n+1} - \mathrm{i} \mu_{\text{c}} \big[ \check{X}_{300} \,, \check{g}_{n+1} \big] = R_n \,,
\end{equation}
where we defined
\begin{equation}
R_n = \big[ \delta \Pi \,, \check{g}_n \big] - \frac{1}{2} \sum_{i=1}^{n} \big\{ \check{g}_i \,, \check{g}_{n+1-i} \big\} + r_n \check{g}_n \,.
\end{equation}
Splitting~$\check{g}_{n+1}$ into parts commuting~($\check{g}_{n+1}^{-}$) and anticommuting~($\check{g}_{n+1}^{+}$) with the matrix~$\check{X}_{300}$ we obtain, again using the normalization condition for~$\check{g}_0$ and introducing the commuting and anticommuting with~$\check{X}_{300}$ parts of~${\bar{R}_n = \bar{R}_n^{-} + \bar{R}_n^{+}}$ with ${ \big[ \bar{R}_n^{-} \,, \check{X}_{300} \big] = 0 = \big\{ \bar{R}_n^{+} \,, \check{X}_{300} \big\} }$, where ${\bar{R}_n = \zeta_{\text{c}}^{-1} \check{g}_{0} R_n / 2}$
\begin{align}
\check{g}_{n+1}^{-} &= \bar{R}_n^{-} \,, \\
\check{g}_{n+1}^{+} &= \frac{1 + \mathrm{i} \mu_{\text{c}} \zeta_{\text{c}}^{-1} \check{g}_{0} \check{X}_{300}}{1 + \zeta_{\text{c}}^{-2} \mu_{\text{c}}^2} \bar{R}_n^{+} \,.
\end{align}

Using the simple relations ${\mathcal{M}^{-} = 2^{-1} \big\{ \mathcal{M} \,, \check{X} \big\} \cdot \check{X}}$ and ${\mathcal{M}^{+} = 2^{-1} \big[ \mathcal{M} \,, \check{X} \big] \cdot \check{X} }$ for a matrix ${\mathcal{M} = \mathcal{M}^{-} + \mathcal{M}^{+}}$ with ${ \big[ \mathcal{M}^{-} \,, \check{X} \big] = 0 = \big\{ \mathcal{M}^{+} \,, \check{X} \big\} }$ we easily find, subsequently, all the corrections~$\check{g}_{n+1}$. For example, the correction~$\check{g}_1$ is found, setting~${n = 0}$, to be
\begin{align}
\check{g}_{1}^{-} &= - \mathrm{i} \frac{\omega \delta\Delta}{\zeta_{\text{c}}^{2}} \check{g}_{0} \cdot \check{\mathrm{X}}_{313} \,, \label{g1-} \\
\check{g}_{1}^{+} &= - \frac{m}{D_{\text{c}}} \check{A}\cdot \check{g}_{0} \cdot \check{\mathrm{X}}_{123} \,, \label{g1+}
\end{align}
where ${D_{\text{c}} = \zeta_{\text{c}}^{2} + \mu_{\text{c}}^{2}}$ and ${\check{A} = \zeta_{\text{c}} \check{g}_{0} + \mathrm{i} \mu_{\text{c}} \check{\mathrm{X}}_{300}}$ with ${\zeta_{\text{c}} = \sqrt{\Delta_{\text{c}}^2 + \omega^2}}$. A correction due to the variation of the parameter~$\mu$ does not arise in this order because the commutator $\big[ \check{\mathrm{X}}_{300} \,, \check{g}_{0} \big]$ is zero.

We concentrate on the derivative term~$\check{g}_{3,k}$ only and, calculating the third order correction and keeping in eye the gradient term only, we obtain
\begin{equation}
\check{g}_{3,k} = v_{x}^2 \frac{\zeta_{\text{c}} (\zeta_{\text{c}}^2 - 3 \mu_{\text{c}}^2)}{4 D_{\text{c}}^3} \partial_x^2 m \cdot \check{X}_{123} \,.
\label{eqn:GradientTerm_Appendix}
\end{equation}

\section{Coefficients in the Ginzburg--Landau-like equation}
\label{sec:Coefficients}

The Ginzburg--Landau-like equation~(\ref{eqn:TimeIndependentEquationForM}) equation is written, using the dimensionless quantities~${\tilde{m} = m / T_{\text{m}}}$, ${\delta\tilde{\mu} = \delta\mu / T_{\text{m}}}$ and~${\tilde{x} = x / l}$ with~${l = v / T_{\text{m}}}$, in the form
\begin{equation}
- \alpha \partial^2_{\tilde{x}} \tilde{m} + \big( \gamma \delta\tilde{\mu} + \beta \tilde{m}^2 \big) \tilde{m} = 0 \,,
\end{equation}
where we defined the coefficients
\begin{align}
\alpha &= 2 \pi T \sum_{\omega = 0}^{\infty} \big\langle 8^{-1} \zeta_{\text{c}} D_{\text{c}}^{-3} \big( \zeta_{\text{c}}^2 - 3 \mu_{\text{c}}^2 \big) \cdot \big( 1 + \cos(2 \phi) \cos(2 \theta) \big) \big\rangle \,, \label{eqn:AlphaTemperature} \\
\beta &= \frac{T_{\text{m}}^2}{2} \left( 2 \pi T \sum_{\omega = 0}^{\infty} \big\langle \zeta_{\text{c}} D_{\text{c}}^{-3} \big( \zeta_{\text{c}}^2 - 3 \mu_{\text{c}}^2 \big) \big\rangle - A_1^2 A_0^{-1} \right) \,, \label{eqn:BetaTemperature}\\
\gamma &= 2 \pi T T_{\text{m}} \sum_{\omega = 0}^{\infty} \big\langle 2 D_{\text{c}}^{-2} \mu_{\text{c}} \zeta_{\text{c}} \big\rangle \,, \label{eqn:GammaTemperature} \\
\intertext{with}
A_0 &= 2 \pi T \sum_{\omega = 0}^{\infty} \zeta_{\text{c}}^{-3} \,, \\
A_1 &= 2 \pi T \sum_{\omega = 0}^{\infty} \big\langle D_{\text{c}}^{-2} \zeta_{\text{c}}^{-1} (\zeta_{\text{c}}^{2} - \mu_{\text{c}}^{2}) \big\rangle \,,
\end{align}
and~$\theta$ is the angle between the Fermi velocity~$\mathbf{v}$ and the $x$\nobreakdash-axis.

In the limit of low temperatures~${T \ll \min (\Delta_{\text{c}}, \mu_{\text{c}})}$ the coefficients can be represented in the form
\begin{align}
\alpha &= - \big( 4 \Delta_{\text{c}}\big)^{-2} \big\langle \rho_{\text{c}}^{-4} \big( 3 s_{\text{c}} \rho_{\text{c}}^{-1} \ln (s_{\text{c}} + \rho_{\text{c}}) + s_{\text{c}}^2 - 2 \big) \notag \\
&\times \big( 1 + \cos(2 \phi) \cos(2 \theta) \big) \big\rangle \,, \label{eqn:AlphaLowTemperature} \\
\beta &= T_{\text{m}}^2 \big( 2 \Delta_{\text{c}} \big)^{-2} \big\langle \rho_{\text{c}}^{-4} \big( 2 - s_{\text{c}}^2 - 3 s_{\text{c}} \rho_{\text{c}}^{-1} \ln(s_{\text{c}} + \rho_{\text{c}}) \big) - 2 \tilde{A}_1^2 \tilde{A}_0^{-1} \big\rangle \,, \label{eqn:BetaLowTemperature}\\
\gamma &= T_{\text{m}} \Delta_{\text{c}}^{-1} \big\langle \rho_{\text{c}}^{-3} \big( s_{\text{c}} \rho_{\text{c}} + \ln (s_{\text{c}} + \rho_{\text{c}}) \big) \big\rangle \,, \label{eqn:GammaLowTemperature} \\
\intertext{with}
\tilde{A}_0 &= 1 \,, \\
\tilde{A}_1 &= \big\langle \rho_{\text{c}}^{-2} \big( 1 - s_{\text{c}} \rho_{\text{c}}^{-1} \ln (s_{\text{c}} + \rho_{\text{c}}) \big) \big\rangle \,,
\end{align}
where~${s_{\text{c}} = \mu_{\text{c}} / \Delta_{\text{c}}}$ and~${\rho_{\text{c}} = \sqrt{s_{\text{c}} + 1}}$.

The time-dependent equation for~$m$, Eq.~\ref{eqn:TimeDependentEquationForM}, is
\begin{equation}
- \alpha \partial^2_{\tilde{x}} \tilde{m} + \big( \gamma \delta\tilde{\mu} + \beta \tilde{m}^2 \big) \tilde{m} = - \varsigma \partial_{\tilde{t}} \tilde{m} \,,
\end{equation}
where ${\tilde{t} = T_{\text{m}} t}$ and the coefficient~$\varsigma$ is
\begin{equation}
\varsigma = 2 \pi T T_{\text{m}} \sum_{\omega = 0}^{\infty} \big\langle 2^{-1} \zeta_{\text{c}}^{-1} D_{\text{c}}^{-2} \omega \big( \zeta_{\text{c}}^2 - \mu_{\text{c}}^2 \big) \big\rangle \,,
\label{eqn:VarSigmaArbitraryTemperatures}
\end{equation}
which in the limit of low temperatures reads
\begin{equation}
\varsigma = 2^{-1} T_{\text{m}} \Delta_{\text{c}}^{-1} \big\langle \big( 1 + s_{\text{c}}^2 \big)^{-1} \big\rangle \,.
\label{eqn:VarSigmaLowTemperatures}
\end{equation}

%\bibliography{Bibliography}

%

\end{document}